# Interplay between grain boundary grooving, stress, and dealloying in the agglomeration of NiSi$_{1-x}$Ge$_x$ films

**H. B. Yao, M. Bouville, and D. Z. Chi,**[*]
*Institute of Materials Research and Engineering, 3 Research Link, Singapore 117602*

**H. P. Sun and X. Q. Pan**
*Department of Materials Science and Engineering, University of Michigan, Ann Arbor, MI 48109-2136, U.S.A.*

**D. J. Srolovitz**
*Department of Mechanical and Aerospace Engineering, Princeton University, Princeton NJ, 08540, U.S.A.*

**D. Mangelinck**
*L2MP-CNRS, Case 151, Faculté de saint Jérôme, Université Paul Cézanne, 13397 Marseille cedex 20, France*

Germanosilicides, especially those formed on compressive substrates, are less stable than silicides against agglomeration. By studying the solid-state reaction of Ni thin film on strained Si$_{0.8}$Ge$_{0.2}$(001), we show that nickel germanosilicide is different from nickel silicide and nickel germanide in several respects: the grains are smaller and faceted, the groove angle is sharper, and dealloying takes place. The germanium out-diffusion creates a stress in the film which favors grooving and agglomeration.

---

[*] **Electronic mail:** dz-chi@imre.a-star.edu.sg

Si$_{1-x}$Ge$_x$ alloys are under consideration for a wide variety of high performance electronic and optoelectronic devices, including advanced metal–oxide–semiconductor field-effect transistors (MOSFETs), where the enhanced carrier mobility of Ge (channel) and the reduced contact resistance (source and drain) are particularly important. Reliable low-resistivity electrical contacts to the Ge-containing regions can be formed using the same process as for self-aligned silicides. NiSi$_{1-x}$Ge$_x$ is the contact of choice in such applications due to its low-resistivity and the low processing temperature needed for its formation. Interfacial reactions of Ni with Si$_{1-x}$Ge$_x$ have been studied in the context of ohmic contacts and Schottky barriers for infrared detectors.[1-4] The downside of metal germanosilicides is their lower stability against agglomeration than the corresponding silicides.[5-8] Moreover, germanosilicide films formed on compressive substrates are less stable than those formed on relaxed substrates.[5-7] There is a tendency for increased Ge content of the Si$_{1-x}$Ge$_x$ alloy at the triple junction, i.e. the intersection of the semiconductor/germanosilicide interface and the grain boundary.[1, 5, 7-9] In this letter, we examine the correlations between agglomeration, composition inhomogeneities and interface morphology in NiSi$_{1-x}$Ge$_x$ films[10] formed on Si$_{0.8}$Ge$_{0.2}$. We demonstrate that germanosilicides exhibit a fundamentally different agglomeration mechanism than the corresponding silicide and germanide films.

Nickel films (15 nm thick) were DC magnetron sputtered (base pressure $< 5 \times 10^{-7}$ torr) onto a 73 nm-thick strained Si$_{0.8}$Ge$_{0.2}$(001). Rapid thermal annealing (RTA) in a N$_2$ ambient for 20 s between 300 and 800 °C was employed to form the nickel germanosilicide. X-ray diffraction (not shown) indicates that only NiSi$_{1-x}$Ge$_x$ was formed during RTA at 400–800 °C. At 300 °C both Ni$_2$Si$_{1-x}$Ge$_x$ and NiSi$_{1-x}$Ge$_x$ were found.

Figure 1(a) is a high-resolution cross-sectional transmission electron micrograph (HRXTEM) of NiSi$_{1-x}$Ge$_x$/Si$_{0.8}$Ge$_{0.2}$ after RTA at 500 °C. A well-defined groove forms at the (triple) junction of the NiSi$_{1-x}$Ge$_x$ grain boundary and the NiSi$_{1-x}$Ge$_x$/Si$_{1-x}$Ge$_x$ interface (the grain boundary groove at the free surface is very shallow). Similarly, grooves form at grain boundaries in NiSi on Si and NiGe on Ge, as shown in Figs. 1(b) and 1(c). The grain boundary groove angle $\theta$ (defined in Fig. 1) in NiSi$_{1-x}$Ge$_x$ is considerably smaller than in NiSi and NiGe (the latter two are of similar magnitude).

Figure 2(a) shows a cross-sectional transmission electron micrograph of NiSi$_{1-x}$Ge$_x$ formed on compressive Si$_{0.8}$Ge$_{0.2}$ at a lower magnification than that in Fig. 1(a). The NiSi$_{1-x}$Ge$_x$/Si$_{1-x}$Ge$_x$ interface in this figure consists of a series of nearly flat sections, joined at relatively sharp corners. A large section of this interface appears to remain parallel to the initial flat (001) surface of the Si$_{1-x}$Ge$_x$ substrate. Although the possible importance of this phenomenon has not previously been emphasized, polygonal NiSi$_{1-x}$Ge$_x$ interface morphologies can also be observed in earlier studies.[5, 6, 9, 11] On the other hand, the NiSi/Si and NiGe/Ge interfaces appear to be much smoother curves (i.e., of nearly constant curvature), as shown by Figs. 2(b) and 2(c). Constant curvature interfaces meeting at grain boundary grooves suggest that capillary equilibrium was established over the entire interface (assuming isotropic incoherent interface energies).

Sheet resistance measurements (not shown) indicate that the resistivity of the NiSi$_{1-x}$Ge$_x$ increasing with RTA temperature above 500°C. Since the x-ray diffraction data indicate that these films contained only one phase, the increasing sheet resistance must be associated with agglomeration (rather than the formation of a high resistivity phase). This interpretation is supported by the plan view scanning electron micrographs (SEM) of Figs.

3(a) and 3(b) which show that little $NiSi_{1-x}Ge_x$ agglomeration has occurred at 500°C and substantial agglomeration has occurred at 650°C.

Comparison of the sheet resistance data for $NiSi_{1-x}Ge_x$ formed on strained $Si_{0.8}Ge_{0.2}$ with that of NiSi on Si suggests that agglomeration begins at temperatures below 500°C in the germanosilicide whereas NiSi films (same thickness and annealing time, 20 s) show little agglomeration to temperatures 100°C higher. This is supported by the SEM micrographs of the $NiSi_{1-x}Ge_x$ and NiSi agglomeration morphologies in Fig. 3. These micrographs also show that the gaps in the $NiSi_{1-x}Ge_x$ are uniformly spaced with a separation comparable with the grain size, while in the NiSi, the hole spacing is non-uniform and several times larger than the grain size. NiGe films on Ge, Fig. 2(d), exhibit a morphology very similar to that of NiSi on Si at a similar degree of agglomeration, Fig. 2(c). This is surprising, since one would expect that the behavior of the $NiSi_{1-x}Ge_x$ should be intermediate between those of NiGe and NiSi.

The results presented above, taken *in toto*, suggest that the grooving, interface morphology, and agglomeration properties of $NiSi_{1-x}Ge_x$ are quite different from those of NiSi and NiGe. Clearly, a major difference between them is that the germanosilicide is a solid solution alloy. As such, the thermodynamics of the system is strongly influenced by alloy thermodynamics rather than just interface energies (the normal basis for discussions of agglomeration). In fact, it may be that the variation of composition and its influence on the thermodynamics and kinetics of the system is responsible for the differences between the $NiSi_{1-x}Ge_x$ and its terminal phases NiSi and NiGe. To this end, we employed energy dispersive x-ray spectroscopy (TEM-EDX) to determine the relative concentrations of Si and Ge at different locations in the microstructure (labeled S1–S8 in Fig. 1(a)).

Each EDX measurement averages over a cylinder of material of approximate diameter 5 nm. The results, shown in Fig. 1(d), suggest that the Ge content of the $Si_{1-x}Ge_x$ alloy and the $NiSi_{1-x}Ge_x$ film are indistinguishable at all locations in these two materials that are not near the grain boundary (spots S1, S4, S5, S7, and S8). Interestingly, the Ge concentration is reduced by approximately a factor of three near the grain boundary in the $NiSi_{1-x}Ge_x$ (spot S2). The Ge concentration in the $Si_{1-x}Ge_x$ alloy immediately below the triple junction (spot S3) is increased by more than a factor of two. Similar observations were made in the Ti–$Si_{1-x}Ge_x$ system.[12] This suggests that the $NiSi_{1-x}Ge_x$ film was formed sufficiently rapidly that there was little time for Si/Ge redistribution and that the Ge depleted from the $NiSi_{1-x}Ge_x$ flowed down the grain boundary toward the triple junction with $Si_{1-x}Ge_x$. The triple junction moves up, trapping the high concentration of Ge along its track.

In the Ni–Si–Ge phase diagram,[8] the tie-lines connect regions with composition near $NiSi_{0.8}Ge_{0.2}$ with regions with compositions near $Si_{0.3}Ge_{0.7}$ at 600 °C. Therefore, as the film/substrate systems evolves, the germanosilicide will expel germanium into the silicon-germanium. Since Ge diffusion is faster along the $NiSi_{1-x}Ge_x$ grain boundary than within a $NiSi_{1-x}Ge_x$ grain the grain boundary is the fastest path for Ge egress. This explains why the region near the grain boundary in the germanosilicide is Ge-poor compared with the rest of the germanosilicide film (see Fig. 1). Since the expelled Ge enters the $Si_{1-x}Ge_x$ directly under the germanosilicide grain boundary, this region should be richer in Ge than the rest of the substrate, as also seen in Fig. 1.

The morphology of the grain boundary groove in $NiSi_{1-x}Ge_x$ is quite different from those in NiGe and NiSi. In the latter cases, the film/substrate interfaces are uniformly curved while the groove has more abrupt features in $NiSi_{1-x}Ge_x$ (see Fig. 2). This is consistent with the much faster grooving seen in NiSi$_{1-x}$Ge$_x$, i.e. when grooving is fast there is not sufficient time available to diffusionally smooth the interfaces. We also note that the apparent groove angle $\theta$ is significantly smaller in NiSi$_{1-x}$Ge$_x$ than in NiSi or NiGe (see Fig. 1). Since NiSi$_{1-x}$Ge$_x$ (Si$_{1-x}$Ge$_x$) are continuous solid solutions between NiSi and NiGe (Si and Ge), we expect that the grain boundary and interface energies are also smooth interpolations between these limits. Therefore, it is doubtful that this large angle change is associated with differences in the interface/grain boundary energies. Rather, we suggest that the morphology observed during the grain boundary grooving in NiSi$_{1-x}$Ge$_x$ is kinetically determined (i.e. the triple junction angle *observed at this scale* is not the equilibrium angle). While Ge diffusion can change the compositions of the two phases, it is only transport of the Ni that controls morphology change (we note that the solubilities of Ni in Si$_{1-x}$Ge$_x$ and NiSi$_{1-x}$Ge$_x$ are negligible).

The difference in groove morphology between the NiSi, NiGe, and NiSi$_{1-x}$Ge$_x$ is likely associated with the fact that of these three, only in NiSi$_{1-x}$Ge$_x$ does the composition vary. Germanium diffuses down the NiSi$_{1-x}$Ge$_x$ grain boundary, as shown in Fig. 4(a), at the same time as the film grooves. This leads to a depletion of Ge and a tensile stress in the film (the atomic volume of nickel silicide is 10% lower than that of nickel germanide) as shown in Fig. 4(b). Conversely the substrate is under compression just below the triple junction since the lattice parameter of Si$_{1-x}$Ge$_x$ increases with Ge content. Formation of new NiSi$_{1-x}$Ge$_x$ at the grain boundary in the film would relieve the tensile stress in the film. This requires short-range Ni diffusion along the grain boundary, Fig. 4(c). (Note, there will also be diffusion of Ni along the interface to remove gradients in interface curvature.) This should occur readily given the fast out-diffusion of Ge and in-diffusion of Si along the grain boundary. Hence, the fast grooving can ultimately be traced to the grain boundary diffusion of Ge out of the NiSi$_{1-x}$Ge$_x$ film. This mechanism does not replace the "usual" grooving mechanism observed in NiSi and NiGe.[13] The two effects combine and lead to faster agglomeration in nickel germanosilicide than in the silicide or germanide. If the grain boundary and interface energies were small the "usual" grooving mechanism would be very slow but the mechanism presented would still work as it does not depend on these energies.

In summary, we demonstrated that the agglomeration of nickel germanosilicide is fundamentally different from that in NiSi and NiGe: it occurs more quickly, it leads to very different grooving morphologies, and composition changes. Of particular importance is the coupling between the change in alloy composition that occurs through fast grain boundary diffusion and the acceleration of grooving associated with inhomogeneous compositional stress effects. The fast grooving in NiSi$_{1-x}$Ge$_x$ leads to groove morphologies that are much farther from equilibrium than in NiSi or NiGe (non-uniform interface curvature). The composition changes that occur in NiSi$_{1-x}$Ge$_x$ are the cause of the widely observed acceleration of agglomeration in this system.

# References

1. S.-L. Zhang, Microelectron. Eng. **70**, 174 (2003).
2. J. Seger, S.-L. Zhang, D. Mangelinck, and H. H. Radamson, Appl. Phys. Lett. **81**, 1978 (2002).
3. T. H. Yang, G. L. Luo, E. Y. Chang, T. Y. Yang, H. C. Tseng, and C. Y. Chang, IEEE Electron Dev. Lett. **24**, 544 (2003).
4. Jian-Shing Luo, Wen-Tai Lin, C. Y. Chang, and W. C. Tsai, Mater. Chem. Phys. **54**, 160 (1998); J. Appl. Phys. **82**, 3621 (1997).
5. J. Seger, T. Jarmar, Z.-B. Zhang, H. H. Radamson, F. Ericson, U. Smith, and S.-L. Zhang, J. Appl. Phys. **96** 1919–1928 (2004).
6. Q. T. Zhao, D. Buca, S. Lenk, R. Loo, M. Caymax, and S. Mantl, Microelectron. Eng. **76**, 285 (2004).
7. Y.-W. Ok, S.-H. Kim, Y.-J. Song, K.-H. Shim, and T.-Y. Seong, Semicond. Sci. Technol. **19**, 285–290 (2004).
8. T. Jarmar, J. Seger, F. Ericson, D. Mangelinck, U. Smith, and S.-L. Zhang, J. Appl. Phys. **92**, 7193 (2002).
9. H. B. Yao, S. Tripathy, and D. Z. Chi, Electrochem. Solid-State Lett. **7**, G323 (2004).
10. We use the notations $NiSi_{1-x}Ge_x$ and $Si_{1-x}Ge_x$ as labels for the nickel germanosilicide and the solid solution silicon germanium alloy. The subscript x is not used to imply a specific composition of either phase or that the two phases have the same germanium content.
11. L. J. Chen, J. B. Lai, and C. S. Lee, Micron **33**, 535–541 (2002).
12. J. B. Lai and L. J. Chen, J. Appl. Phys. **86**, 1340 (1999).
13. T. P. Nolan, R. Sinclair, and R. Beyers, J. Appl. Phys. **71**, 720 (1992).

**Figure Captions:**

**Figure 1.** HRXTEM images of films annealed for 20 s: (a) NiSi$_{0.8}$Ge$_{0.2}$ at 500 °C, (b) NiSi at 650 °C, and (c) NiGe at 500 °C. Figure (d) shows TEM-EDX data obtained from the regions S1–S8 indicated in Fig. (a); the spectra were analyzed to estimate the Ge/Si peak height ratio.

**Figure 2.** HRXTEM images of (a) a NiSi$_{1-x}$Ge$_x$ film formed on strained Si$_{0.8}$Ge$_{0.2}$ with RTA for 60 s at 550 °C, (b) a NiSi film on Si with RTA for 60 s at 750 °C, and (c) a NiGe film on Ge with RTA for 20 s at 500 °C.

**Figure 3.** Plan view SEM images (a) of a NiSi$_{0.8}$Ge$_{0.2}$ film annealed at 500 °C and (b) 650 °C, (c) a NiSi film annealed at 600 °C and (d) a NiGe film annealed at 500 °C. Light colored areas correspond to the film and dark areas to the exposed substrate.

**Figure 4.** Schematic of the grooving mechanism in nickel germanosilicide. Figure (a) shows the Ge out-diffusion, Fig. (b) the stress state in the system, and Fig. (c) the in-diffusion of Ni and Si.

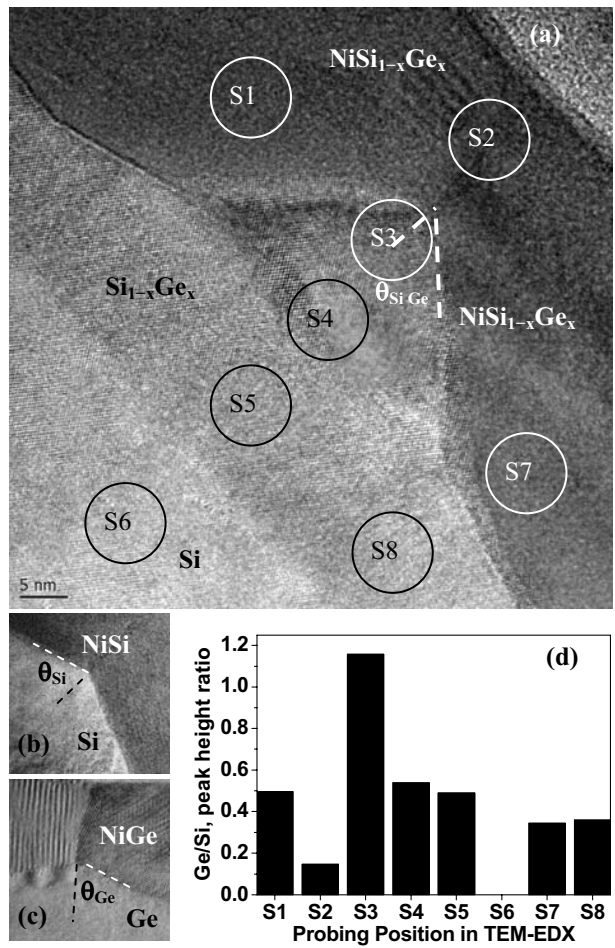

**Figure 1**
**H.B. Yao** *et al*

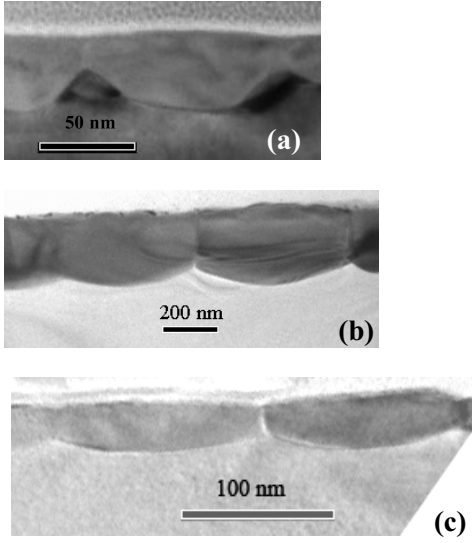

**Figure 2**
**H.B. Yao** *et al*

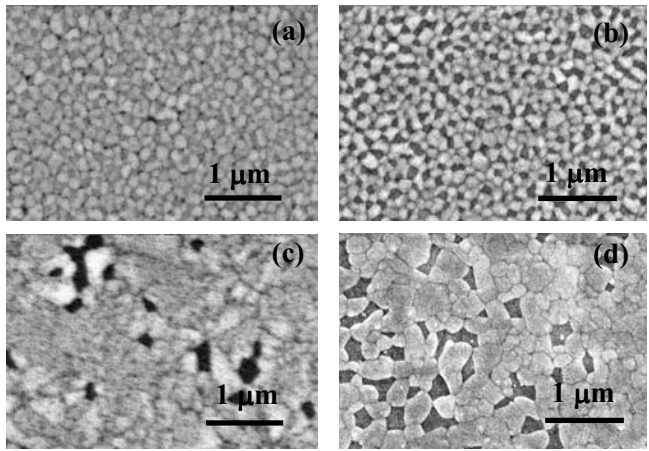

**Figure 3**
**H.B. Yao** *et al*

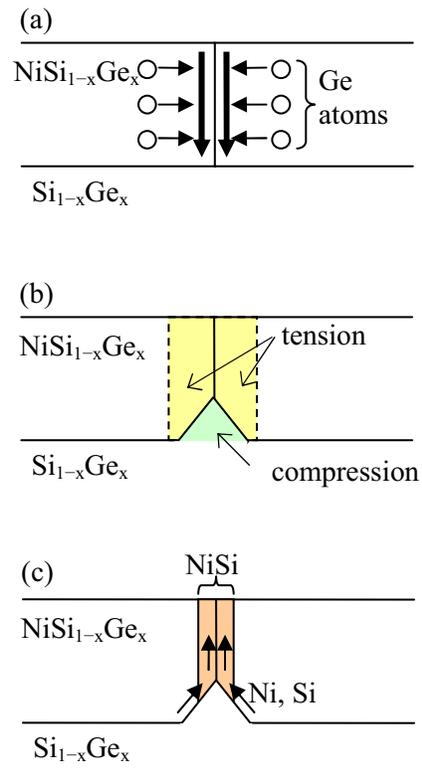

**Figure 4**
**H.B. Yao** *et al*